\documentclass[reprint,aps,prx,twocolumn,
floatfix,tightenlines,
longbibliography,balancelastpage,
]{revtex4-2}
\usepackage{amsfonts}
\usepackage{color}
\usepackage{amsmath}    
\usepackage{epsfig}
\usepackage{graphicx}
\usepackage{dcolumn}
\usepackage{bm}
\usepackage[breaklinks,colorlinks = true,linkcolor = blue,urlcolor=blue,citecolor=blue]{hyperref}
\usepackage[english]{babel}
\babeladjust{autoload.bcp47 = on}
\definecolor{dgreen}{rgb}{0,0.7,0}

\newcommand{\da}{\mathop{}\!\mathrm{d}}

\newcommand{\mc}[1]{\mathcal{#1}}

\newcommand{\mf}[1]{\mathfrak{#1}} 
\newcommand{\mt}[1]{\mathtt{#1}} 
\newcommand{\mr}[1]{\mathrm{#1}}
\usepackage{tikz}
\usetikzlibrary{arrows.meta, positioning}

\begin{document}
\title{Dean-Kawasaki fluctuating hydrodynamics for backscattering hard rods}
\author{Mrinal Jyoti Powdel}
\email{mrinal.jyoti@icts.res.in}
\affiliation{International Centre for Theoretical Sciences, Tata Institute of Fundamental Research, Bengaluru 560089, India}

\begin{abstract}
We study a system of backscattering hard rods in one dimension. Contrary to 
the usual ballistic hard rods, these hard rods flip the sign of their 
velocities with a rate $\gamma$. This leads to the decay of the odd 
moments of velocity while preserving the even moments: the number of 
conserved quantities in the system becomes half. The introduction of 
the flipping rate $\gamma$ is a integrability-breaking 
perturbation and this leads to a change in the transport properties in the 
system. We show using a Dean-Kawasaki fluctuating hydrodynamic formulation that the unequal space-time correlation of the normal mode phase space densities attains a diffusive form at late times. Also, we show that for $t \gg 1/\gamma$, 
the two-time density density correlation of mass densities spreads in a diffusive manner, 
and for $t \ll 1/\gamma$, the correlation spreads ballistically, for a background state given by the Boltzmann distribution. Our results present a elegant framework to study systems where integrability is broken by a stochastic noise.
\end{abstract}

\maketitle
Over the past decade, the framework of generalised hydrodynamics (GHD) has been notably successful in providing
a reliable coarse-grained theory for out-of-equilibrium dynamics in integrable systems  \cite{Castro_Doyon_2016, Bertini_Jacopo_2016, doyon_perspective, Science_Neel_2021, Alba_2021, Doyon_GHD_Toda_2019}.
Interest in such systems stemmed in large part from the seminal experimental realization of trapped 1D cold atoms that failed to equilibrate over long times due to integrable dynamics \cite{kinoshita2006quantum, Quantum_Newton_Cradle_scipost}.
Such realizations of integrable systems is, however, not an easy feat, since integrability is a fine-tuned property and an integrable system is
highly sensitive to inhomogeneities and external perturbations. This explains
why most physical systems are either non-integrable or are only approximately
integrable \cite{Bastianello_2021}. In the presence of integrability breaking perturbations, it is hence
naturally expected that in the long-time limit, the framework of GHD reduces to 
the conventional hydrodynamics. The understanding of the precise mechanism that 
leads to this transition is a very interesting question, one that is also experimentally 
tractable, owing to the tunability of integrability breaking perturbations \cite{tang_thermalization_2018}. 
Various implementations \cite{jung_spin_2007, znidaric_coexistence_2013, karrasch_spin_2015, 
biella_energy_2016, steinigeweg_scaling_2014-1, surace_weak_2023, akemann_two_2025, 
bulchandani_onset_2022} of integrability breaking mechanisms have confirmed that
a breakdown of integrability leads to an onset of chaoticity through a change in the 
transport properties of the system. A proper theoretical understanding of the underlying 
mechanism, hence, requires an in-depth analysis of various integrable models under different
integrability breaking mechanisms.

Among classical integrable systems, a very good staging ground to understand the 
consequence of integrability breaking is a one-dimensional system of 
hard rods \cite{Tonks_1936}. A hard-rod system is simply a collection of finitely-sized
segments (or rods) that interact with one another through a hard core
potential. The existence of only two-body elastic collisions renders the system integrable and preserves the initial configuration of the hard rod velocities at all times. Owing to its analytical tractability and easy numerical implementation, experiments on integrability breaking can be conveniently carried out in this system. 
For example, it has been shown \cite{Cao_Bulchandani2018} that when integrability is 
broken in this system by means of confinement in a harmonic trap, it displays a 
transient chaotic regime in its dynamics before ultimately failing to thermalize 
at long times. Integrability can also be broken by modifying the dynamics of individual hard rods in a way that the initial configuration of hard rod velocities is not preserved at all times. 
\begin{figure}
    \centering
    \includegraphics[width=0.9\linewidth]{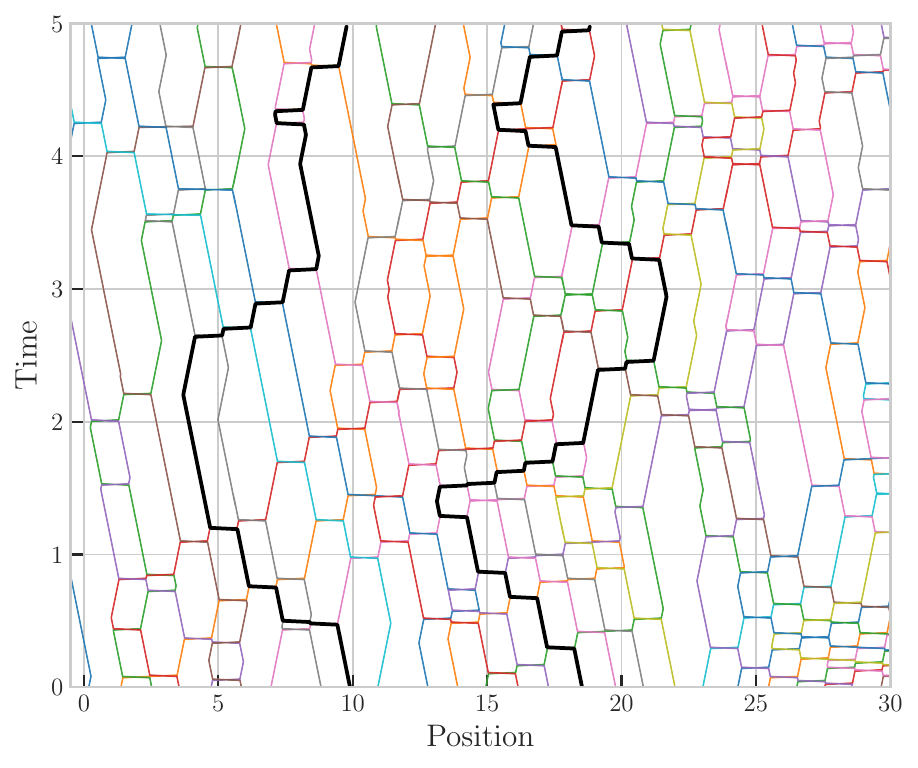}
    \caption{Trajectories of quasiparticles (velocity tracers) in a system of backscattering hard rods with $\gamma = 1$. Two such trajectories have been colored black to aid the reader. The flat lines represent jumps indicating collision events. The change in the direction of motion of a quasiparticle is due to flipping events. }
    \label{fig:trajectories}
\end{figure}

One such mechanism is to introduce a stochastic component in the hard rod dynamics such that they flip the sign of their velocities at random times. 
Previously studied in \cite{piqueresbackscattering2023}, this modification gives rise to backscattering events in the dynamics of hard rods in addition to the usual inter-particle scattering due to collisions. 
Hence, even though the configuration of individual speeds (or unsigned momenta) remains preserved at all times, the configuration of velocities (or signed momenta) does not. 
As an immediate effect of this, the number of conserved charges gets reduced by half: only the even moments of velocity survive while the odd moments decay. 

In this work, we study the departure from integrability in a system of backscattering hard rods.
We develop a Dean-Kawasaki fluctuating hydrodynamic 
formulation for this system and use it to compute unequal space-time correlations in the system. We find that the correlation of the normal mode phase space densities attains a diffusive form at late times and admits an effective diffusion constant. Similarly, for mass densities, the correlation displays a ballistic space time scaling for small times and diffusive space-time scaling for late times, and hence admits a crossover from ballistic transport to diffusive transport.
Our results are consistent with the ones obtained in Ref.~\cite{piqueresbackscattering2023}.  
This, in turn,
demonstrates the wider applicability of a Dean-Kawasaki fluctuating hydrodynamic description in the analysis of an integrable system where individual quasiparticle dynamics is affected by a stochastic noise.

We begin by considering a system of 
$N$ hard rods, each having length $a$ and unit mass
moving in one dimension. Each rod interacts with other rods through two-body elastic collisions and the system at any time, $t$ is fully described by the set  
$\{\mt{x}_i (t),w_i (t),\sigma_i(t) \}$ where $\mt{x}_i (t)$ is the position, $w_i (t)$ is the speed and $\sigma_i (t)$ is the direction of propagation of the $i^\text{th}$ hard rod at time, $t$. In the context of GHD, a system of hard rods is equivalently described in terms of its velocity tracers or quasiparticles.
These are particles which follow the bare velocities of individual hard rods. 
While a collision between two hard rods is depicted as scattering events in their trajectories with scattering length $a$, the same is depicted as `jump' events in quasiparticle trajectories (shown by intermittent plateaus in the quasiparticle trajectories in Fig.~\ref{fig:trajectories})
We would mainly be studying the system from the point of view of quasiparticles. 

A quasiparticle, in a system of ballistic hard rods, moves ballistically with its bare velocity in between collisions such that in the event of a collision from the right, a quasiparticle jumps by a rod length $a$ towards the right and vice versa. 
Needless to say, the direction of motion of a quasiparticle remains the same unless its inherent dynamics changes the sign of its velocity. This is accomplished in this work by adding a stochastic backscattering noise to the quasiparticle dynamics. The resultant modified dynamics is such that any quasiparticle can randomly flip the sign of its velocity with a rate, $\gamma$. The resulting system is a system of backscattering hard rods (BHRs)

Fig.~\ref{fig:trajectories} shows quasiparticle trajectories in a system of BHRs for a particular realization of initial condition. The jump events in quasiparticle trajectories is shown as random sharp horizontal lines or intermittent plateaus. 
At this point, an alternative description of the flipping dynamics of quasiparticles seems convenient. Each backscattering quasiparticle can be said to possess, in addition to $w$ and $\sigma$, a stopwatch that is set to a waiting time, $\mt{t}$ drawn from an exponential distribution, $\gamma e^{-\gamma \mt{t}}$, where $\gamma$ is the flipping rate. As soon as the timer in the stopwatch reaches zero, a left moving quasiparticle ($\sigma = -1$) turns into a right moving one ($\sigma = +1$) and vice-versa, and the stopwatch of the particular quasiparticle is restarted from a new randomly drawn time, $\mt{t}'$.
Fig.~\ref{fig:Schematic of collision dynamics} shows a schematic of the collision dynamics between two backscattering hard rods. In the event of a collision, while two quasiparticles only undergo jumps with no change to their momenta or their waiting times, exchanges of both the momenta and clocks happen in the case of physical hard rods.

\begin{figure}
    \fbox{
    \begin{tikzpicture}[
    x=1cm,y=1cm,
    >=Stealth,
    rod/.style={line width=2.2pt, line cap=round, color=blue!70!black},
    vel/.style={-{Stealth[length=2.4mm,width=1.8mm]}, line width=1pt, color=teal!70!black},
    exch/.style={<->, dashed, line width=1pt, color=orange!85!black},
    clockface/.style={draw=violet!65!black, line width=0.9pt, fill=violet!6},
    hand/.style={line width=0.9pt, color=violet!75!black},
    panel/.style={rounded corners=6pt, draw=gray!45, fill=gray!4},
    title/.style={font=\bfseries\ttfamily\small, fill=white, inner sep=3pt, rounded corners=2pt},
    vellabel/.style={font=\small, text=teal!70!black, fill=white, inner sep=1pt},
    timelabel/.style={font=\small, fill=white, inner sep=1pt},
    exlabel/.style={font=\small, fill=white, inner sep=1.5pt}
]

\filldraw[panel] (-0.20,  1.75) rectangle (5.80,  4.55);
\filldraw[panel] (-0.20, -2.00) rectangle (5.80,  1.40);
\filldraw[panel] (-0.20, -5.15) rectangle (5.80, -2.25);

\node[title] at (2.80, 4.22) {Before Collision};
\node[title] at (2.80, 1.10) {Collision \& Exchange};
\node[title] at (2.80,-2.52) {After Collision};


\draw[rod] (0.35,3.45) -- (1.45,3.45);
\draw[vel] (0.45,3.65) -- (1.65,3.65);
\node[vellabel] at (1.05,3.85) {$v_1$};

\draw[clockface] (0.90,2.65) circle (0.38);
\draw[hand] (0.90,2.65) -- (0.90,2.91);
\draw[hand] (0.90,2.65) -- (1.05,2.74);
\node[timelabel] at (0.90,1.98) {$T_1$};

\draw[rod] (4.10,3.45) -- (5.20,3.45);
\draw[vel] (5.10,3.65) -- (3.90,3.65);
\node[vellabel] at (4.50,3.85) {$v_2$};

\draw[clockface] (4.65,2.65) circle (0.38);
\draw[hand] (4.65,2.65) -- (4.47,2.47);
\draw[hand] (4.65,2.65) -- (4.77,2.53);
\node[timelabel] at (4.65,1.98) {$T_2$};


\draw[rod] (1.75,0.25) -- (2.75,0.25);
\draw[rod] (2.85,0.25) -- (3.85,0.25);
\fill[orange!85!black] (2.80,0.25) circle (0.03);

\draw[exch] (1.95,0.45) -- (3.65,0.45);
\node[exlabel] at (2.80,0.72) {$v_1 \leftrightarrow v_2$};

\draw[clockface] (2.15,-0.35) circle (0.38);
\draw[hand] (2.15,-0.35) -- (2.15,-0.29);
\draw[hand] (2.15,-0.35) -- (2.30,-0.46);
\node[timelabel] at (2.15,-1.05) {$T_1$};

\draw[clockface] (3.45,-0.35) circle (0.38);
\draw[hand] (3.45,-0.35) -- (3.27,-0.73);
\draw[hand] (3.45,-0.35) -- (3.57,-0.67);
\node[timelabel] at (3.45,-1.05) {$T_2$};

\draw[exch] (2.35,-1.35) -- (3.25,-1.35);
\node[exlabel] at (2.80,-1.66) {$T_1 \leftrightarrow T_2$};


\draw[rod] (0.35,-3.45) -- (1.45,-3.45);
\draw[vel] (1.25,-3.25) -- (0.05,-3.25);
\node[vellabel] at (0.65,-3.00) {$v_2$};

\draw[clockface] (0.90,-4.25) circle (0.38);
\draw[hand] (0.90,-4.25) -- (0.72,-4.43);
\draw[hand] (0.90,-4.25) -- (1.02,-4.37);
\node[timelabel] at (0.90,-4.92) {$T_2$};

\draw[rod] (4.10,-3.45) -- (5.20,-3.45);
\draw[vel] (4.30,-3.25) -- (5.50,-3.25);
\node[vellabel] at (4.90,-3.00) {$v_1$};

\draw[clockface] (4.65,-4.25) circle (0.38);
\draw[hand] (4.65,-4.25) -- (4.65,-3.99);
\draw[hand] (4.65,-4.25) -- (4.80,-4.16);
\node[timelabel] at (4.65,-4.92) {$T_1$};

\end{tikzpicture}
}
    \caption{The figure shows a schematic of the collision dynamics of backscattering hard rods. Two rods with velocities and clocks denoted by $(v_1, T_1)$ and $(v_2, T_2)$ approach each other (upper panel), collide (center panel) and exchange both the velocities and clocks (bottom panel). Equivalently, in the quasiparticle description, if one follows a particular velocity, (say $v_1$) the clock ($T_1$) remains attached to the quasiparticle at all times. }
    \label{fig:Schematic of collision dynamics}
\end{figure}

At this stage, it is natural to draw comparisons of this system with interacting run and tumble particles (RTPs). 
In the latter, the interactions among individual RTPs can be modeled by a velocity field in position space which depends on the density of neighboring particles.
\begin{figure*}
    \includegraphics[height = 0.25\linewidth,width=0.99\linewidth]{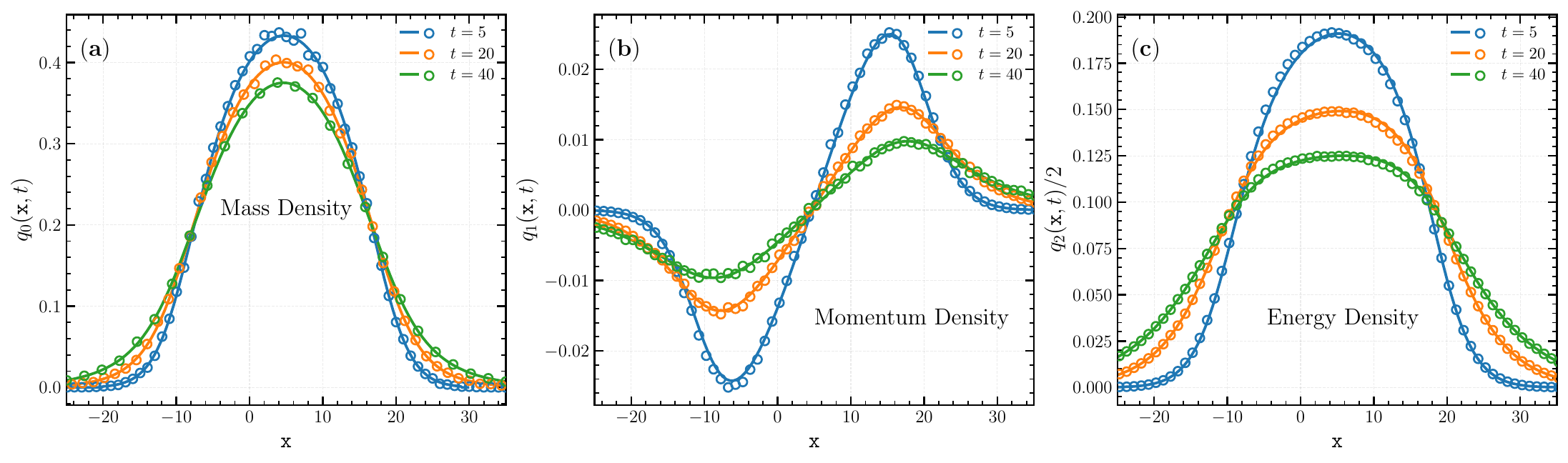}
    \caption{Excellent agreement of the conserved densities computed following the steps outlined in Ref.~\cite{powdel2024conserved}. The figure shows mass density, $q_0 (\mt{x},t)$, momentum density, $q_1 (\mt{x},t)$ and energy density, $q_2 (\mt{x},t)/2$ in a system of backscattering hard rods of length $a=1$ at different times. The conserved densities are defined as $q_\alpha (\mt{x},t) = \sum_\sigma \int \da w ~w \sigma \bar{f}_\sigma (\mt{x},w,t) $. The solid lines represents theory (Appendix \ref{app:conserved_densities}) and the circles represent the results from molecular dynamics. Here, the initial positions and velocities are drawn from a gaussian distribution in Eq.~\eqref{eq:initial_cond_gaussian} with $\sigma_x = 5$ and $T=1$. The numerical results have been obtained after averaging over $5 \times 10^6$ realizations.}
    \label{fig:gaussian_massden_enerden}
\end{figure*}
The resulting problem is a much difficult one and has been explored in the past (see \cite{Tailleur2008RTP}). 
In this work, however, the flipping rate is originally introduced in the quasiparticle dynamics, which is relevant to integrability breaking in GHD, rather than the actual hard rods. Hence, the description of the system naturally admits exchange of clocks between individual hard rods during collision.

Interacting integrable systems such as the system of hard rods are studied within GHD on large space-time scales in terms of the mean phase space densities (PSDs), $\bar{\mf{f}}_\sigma(\mt{x},w,t)$ of the quasiparticles, which is defined for BHRs as the following:
\begin{equation}\label{eq:mainfluctuating_PSD}
\begin{split}
    \bar{\mf{f}}_\sigma(\mt{x},w,t) &= \langle \mf{f}_\sigma (\mt{x},w,t)\rangle \\[3pt]
    \text{where,} \ \ \ 
    \mf{f}_\sigma (\mt{x},w,t) &= \sum_i \mf{f}^{(i)}_\sigma (\mt{x},w,t) \ \ \\[3pt] \text{and}  \ \ \ \mf{f}^{(i)}_\sigma (\mt{x},w,t) &=  \delta_{\sigma_i (t), \sigma}~ \delta (\mt{x}-\mt{x}_i (t))~ \delta(w - w_i) \ \ .
\end{split}
\end{equation}
Here, $\mf{f}_\sigma (\mt{x},w,t)$ is the fluctuating PSD and the angular brackets denote averaging over initial conditions. The index `$i$' is the index of the quasiparticle. It so happens that
the computation of mean PSDs in a system of hard rods
becomes very convenient via the following mapping \cite{LPS_1968,Percus_1969}
\begin{equation}\label{eq:rod_to_pp_map}
\begin{split}
    x_i(t) = x(\mt{x}_i (t)) &= \mt{x}_i (t) - a \sum_{j \neq i} \Theta (\mt{x}_i (t) - \mt{x}_j (t)) , \\ w_i (t) &= w_i (t) \ , \ \sigma_i (t) = \sigma_i (t)
\end{split}
\end{equation}

\begin{figure*}
    \includegraphics[height = 0.32\linewidth,width=0.99\linewidth]{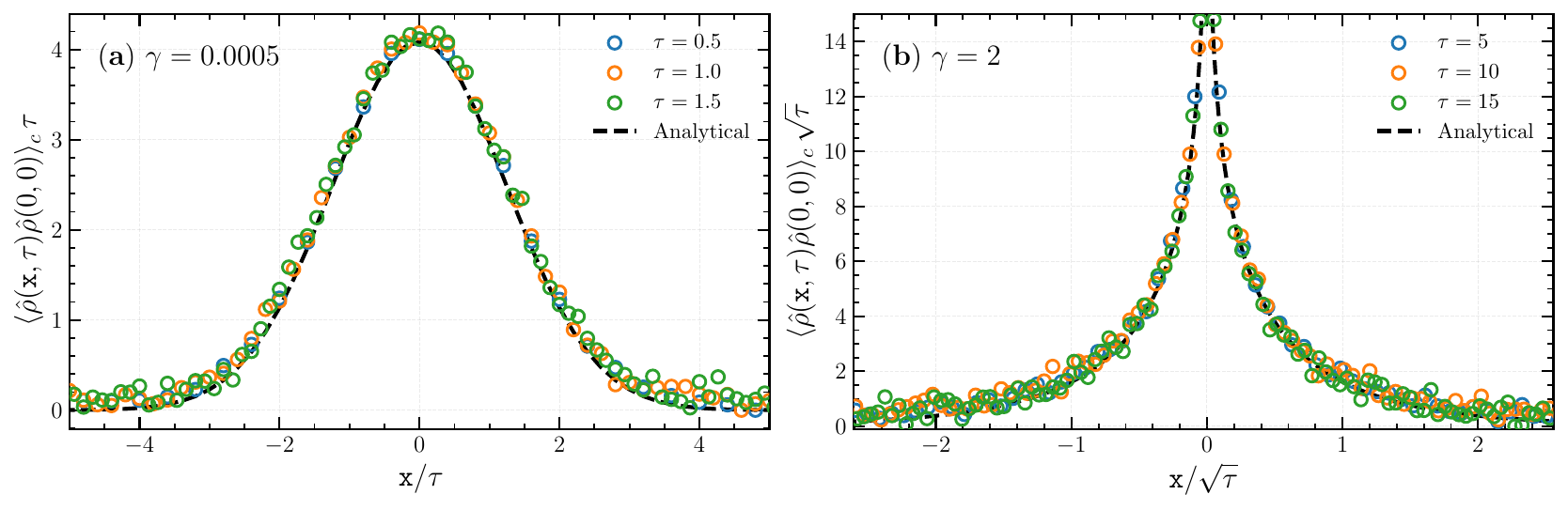}
    \caption{Unequal space-time correlation of mass densities for a hard rod system consisting of $N=400$ backscattering hard rods each of length $a=0.01$ for a) $\gamma = 0.0005$ and b) $\gamma = 2$. Here, we have set $\mt{x'} =0,t'=0$. The black dashed lines represent theory (Eq. ~\eqref{eq:unequal_time_space_time}) and the colored circles represent results from molecular dynamics for different values of $\tau = t-t'$, where we have also subtracted off fluctuations due to finite size effects. Initially, all the hard rods are distributed uniformly in the region $[-10,10]$ and the velocities were drawn from a gaussian distribution with temperature or variance, $T=1$. The boundary conditions imposed are periodic. On the left, we see that for $\tau << \gamma^{-1}$, the correlation has a ballistic space-time scaling. While, on the right, for $\tau>>\gamma^{-1}$, the correlation has a diffusive space-time scaling. The numerical results have been obtained after averaging over $10^6$ realizations.}
    \label{fig:correlation_scaling}
\end{figure*}
\noindent where, by removing the inaccessible space between consecutive hard rods below $\mt{x}_i$, the position, speed and direction of propagation of the $i^{th}$ BHR is mapped to the same for the $i^{th}$ point particle (PP) denoted by $\{x_i (t),w_i, \sigma_i (t)\}$. Here, 
$\Theta (x)$ is the Heaviside Theta function. The procedure to compute various conserved densities analytically for the system of BHRs has been described in brief in Appendix~\ref{app:conserved_densities} following Ref \cite{powdel2024conserved} and the comparison with numerics is provided in Fig.~\ref{fig:gaussian_massden_enerden}. 
As is evident from the figure, there is an excellent agreement shown by analytical form of the conserved densities with numerics.

Understanding how a system responds to fluctuations is one of the main objectives of statistical physics. This is possible by an analysis of the fluctuating PSDs, as opposed to the mean PSDs in which the information about the fluctuations is lost as a result of averaging.
The Dean-Kawasaki equation \cite{DavidSDean_1996} is a stochastic partial differential equation which was originally proposed for a system of interacting Langevin processes interacting via a pairwise potential, and has ever since, found applications in various studies, even beyond systems having a Brownian noise \cite{Tailleur2008RTP, Illien_2025}. The formalism has ultimately given rise to a stochastic density functional theory and is a crossroad between various areas of physics and mathematics. 

To write down the Dean-Kawasaki equation for a system of hard rods affected by a stochastic backscattering noise, we use the inverse 
mapping of Eq.~\eqref{eq:rod_to_pp_map},
\begin{equation}
    \mt{x}_i (t) = x(\mt{x}_i(t)) + a ~ \sum_{j \neq i} \Theta (x(\mt{x}_i(t)) - x(\mt{x}_j (t))) \ \ , 
\end{equation}
which defines the trajectory of the $i^\text{th}$ backscattering hard rod (BHR) at time, $t$ in terms of the positions of the corresponding PPs. Here, $\Theta(x)$ is the heaviside theta function such that $\Theta(x) = 1 \ \text{for} \ x\geq0$ 
and 0 otherwise. In the PP picture, the system corresponds to run-and-tumble particles (RTPs) with hard core interactions. However, since the PPs have no finite extent and the effect of the interaction between two RTPs is a mere exchange of velocities and clocks without the occurence of finite jumps in the trajectories of their velocity tracers, the system can be considered as effectively non-interacting \cite{powdel2024conserved}. An important consequence of this realization is that the trajectory of a quasiparticle in a system of hard rods can be efficiently and conveniently tracked by the trajectory of the corresponding non-interacting RTP.

This turns our focus to evaluating the trajectory of an RTP. In a time interval $\da t$, the displacement of a non-interacting RTP can be described as a Ito-Langevin process: a ballistic displacement for time, $\da t$, with fluctuations arising due to flipping event. Since the event of a velocity flip is a Poisson process, the properties of the noise leading to the flipping event is known. By mapping the RTP trajectory back to the rod picture, and following the steps outlined 
in Appendix ~\ref{app:derivation_fluctuating_hydro}, one obtains the following equation for the fluctuating PSD of BHRs,
\begin{equation}\label{eq:mainfluc_hydrodynamics}
\begin{split}
    &\partial_t \mf{f}_\sigma (\mt{x},w,t) + \partial_\mt{x} {v_\sigma}_{\rm eff} (\mt{x},w,t) \mf{f}_\sigma (\mt{x},w,t)\\[3pt] = &\gamma (\mf{f}_{-\sigma} (\mt{x},w,t) - \mf{f}_{\sigma} (\mt{x},w,t)) +  \sigma \sqrt{\gamma~ \bar{\mf{f}}(\mt{x},w,t)} ~ \xi(\mt{x},w,t)
\end{split} 
\end{equation}
where
\begin{equation}
    {v_\sigma}_{\rm eff} (\mt{x},w,t) = \frac{w \sigma - a \sum_{\sigma'} \int \da \mr{w} ~ \mr{w} \sigma'~ \bar{\mf{f}}_{\sigma'} (\mt{x},\mr{w},t)}{1-a \bar{\rho} (\mt{x},t)}
\end{equation}
is the effective velocity of the quasiparticle at position $\mt{x}$ with bare velocity $w \sigma$ at time, $t$, and the noise $\xi(\mt{x},w,t)$ has a zero mean with the following correlation:
\begin{equation}
        \langle \xi (\mt{x},w,t) ~\xi (\mt{x}',w',t') \rangle_c =  \delta(\mt{x}-\mt{x}') \delta(w-w') \delta(t-t') \ .  
\end{equation}

One must note that in our derivation of Eq.~~\eqref{eq:mainfluc_hydrodynamics}, we have not done a scale by scale coarse graining either in space or time. It is also important to note that the fluctuating PSD, $\mf{f}$ should be seen as a density operator. While Eq.~\eqref{eq:mainfluc_hydrodynamics} superficially resembles the Fokker-Planck equation in Eq.~\eqref{eq:PP_Fokker-Planck}, it is distinguished from the latter because it incorporates the effect of hard core interactions in the effective velocity, ${v_\sigma}_{\rm eff}$ of the quasiparticle and
with the presence of a stochastic noise term on the right hand side signifying the fluctuating nature of the PSDs. Both these terms are sources of non-linearities in the equation.
If one takes an average over initial conditions on both sides of Eq.~\eqref{eq:mainfluc_hydrodynamics}, the noise term vanishes and the resultant equation gives the evolution equation for the mean PSD. The presence of the noise term is crucial in the computation of unequal space-time correlations. 

For instance, one might be interested to compute the unequal space-time correlation of the normal modes in the problem, $\mf{f}^n (\mt{x},w,t) $ which is defined as the following
\begin{equation}
\begin{split}
    \mf{f}^n (\mt{x},w,t) &= \sum_\sigma \mf{f}_\sigma^n (\mt{x},w,t) \ \ , \ \ \text{where,} \\
    \mf{f}_\sigma^n (\mt{x},w,t) &= \frac{\mf{f}_\sigma (\mt{x},w,t)}{(1-a \rho (\mt{x},t))} \ .
\end{split}
\end{equation}
The first step is to linearize Eq.~\eqref{eq:mainfluc_hydrodynamics}. We consider the system in a state, $\mf{f}_\sigma (\mt{x},w,t)$ that is slightly perturbed from a homogeneous stationary state, i.e., $\mf{f}_\sigma (\mt{x},w,t) = \bar{\mf{f}}_0(w)/2 \ + \delta \mf{f}_\sigma (\mt{x},w,t)$,
where $\bar{\mf{f}}_0(w)$ is the full homogeneous mean PSD 
and $\delta \mf{f}_\sigma (\mt{x},w,t)$ denotes a fluctuation in space-time. The linearized version of Eq.\eqref{eq:mainfluc_hydrodynamics} attains a simple form in terms of the normal mode PSDs and can be easily solved in the Fourier domain, where the properties of the noise can be written as,
\begin{equation}\label{eq:noise_property_in_fourier_space}
\begin{split}
    \langle \tilde{\xi} (k,w,t) \rangle &= 0 \ , \\[3pt]
    \langle \tilde{\xi} (k,w,t)~ \tilde{\xi} (k',w',t') \rangle &= \\ 2 \pi~ \delta(k+k') &~\delta(w-w') ~ \delta(t-t') \ .
\end{split}
\end{equation}
For times $t,t' >>\gamma^{-1}$, the correlations are independent of the initial form of the initial fluctuation, $\delta \mf{f}(\mt{x},w,0)$ and are only dependent on the properties of the noise term given in Eq.~\eqref{eq:noise_property_in_fourier_space}. Once the dust is cleared, the full space-time correlation of the normal modes is given in Eq.~\eqref{eq:correlation_normalised} and for large $t-t'$, it gets simplified to the following form (for $t>t'$):
\begin{equation}\label{eq:diffusive_late_time_behaviour}
\begin{split}
    \langle \mf{f}^n (\mt{x}, w, t) &\mf{f}^n (\mt{x}', w', t') \rangle_c \\ = &\frac{\bar{\mf{f}}_0 (w)}{ (1-a \varrho_0)^2} \delta (w - w') \frac{\exp[- \frac{(\mt{x}-\mt{x}')^2}{2 \mc{D}(w) (t-t')}]}{\sqrt{2 \pi \mathcal{D}(w) (t-t')}} \ .
\end{split}
\end{equation}
Thus the space-time correlation of the normal mode PSDs, $\mf{f}^n$ has a diffusive form with an effective diffusion constant given by $\mc{D}(w) = w^2/(\gamma (1-a\varrho_0)^2)$.
One might also be motivated to compute similar unequal space-time correlations for the conserved densities in the system. We show the computation below for mass densities, $q_0 (\mt{x,t}) = \varrho(\mt{x},t)$, for which the following form of
the homogeneous stationary state can be chosen:
\begin{equation}\label{eq:Gibbs_state}
    \bar{\mf{f}}_0 (w) = \varrho_0~ \frac{2}{\sqrt{2 \pi~ T}} \exp\bigg[- \frac{w^2}{2 T}\bigg] \  ,
\end{equation}
which is a constant function of $\mt{x}$ and an even function of velocity such that the odd moments of the velocity,~$v=w \sigma$ are 0.
This is the well-known Boltzmann distribution which describes the density of hard rods having kinetic energy $w^2/2$ when the system is at temperature, $T$. We define $\tau = t-t'$.
At two regimes, $\tau<< \gamma^{-1}$ and $\tau>>\gamma^{-1}$, the calculation is simpler to perform and the steps are outlined in Appendix \ref{app:linearized_fluc} resulting in a final form given by
\begin{subequations}\label{eq:unequal_time_space_time}
    \begin{equation}
    \begin{split}
        &\langle \varrho (\mt{x},t) ~ \varrho (\mt{x}',t') \rangle_c \\[3pt] &=  \frac{e^{-\gamma \tau} \varrho_0 (1-a\varrho_0)^3}{\sqrt{2 \pi T~ \tau^2}} \exp \bigg[ - (1-a\varrho_0)^2~\frac{(\mt{x}-\mt{x}')^2}{2 T \tau^2} \bigg] \\[3pt]
        & \ \ \ \ \ \ \ \ \ \ \ \ \ \ \ \ \ \text{for } \tau << \gamma^{-1} \text{ and,}
    \end{split}
\end{equation}
\begin{equation}
    \begin{split}
        &\langle \varrho (\mt{x},t) ~ \varrho (\mt{x}',t') \rangle_c  \\ &=  \frac{\varrho_0 (1-a\varrho_0)^3 }{\pi} \sqrt{\frac{\gamma}{T\tau}} K_0 \bigg[(1-a\varrho_0)\sqrt{\frac{\gamma}{T \tau}}~|\mt{x}-\mt{x}'|  \bigg]  \\
    & \ \ \ \ \ \ \ \ \ \ \ \ \ \ \ \ \ \text{for } \tau >> \gamma^{-1} \ .
    \end{split}
\end{equation}
\end{subequations}
Here, $K_0$ is the Bessel `K' function of order 0.
As is clearly seen, the unequal space-time correlation of mass densities display different space-time scalings at the two regimes: the space-time scaling is ballistic for smaller times and it is diffusive for large times. The ballistic component of the correlation function dominates for $t-t' << \gamma^{-1}$ and the diffusive component dominates for $t-t' >> \gamma^{-1}$, resulting in a crossover from ballistic transport to diffusive transport of mass in a system of BHRs. One could also compute the correlation for intermediate times, starting from an extensive numerical integration over $w,w'$ of the normal mode correlation given in Eq.~\eqref{eq:correlation_normalised} and using the relation between correlations in normal mode mass densities and physical densities provided in Eq.~\eqref{eq:correlation_relation_normal_unnormal}. However, this is not pursued in this work. The agreement of the molecular dynamics simulation with the ballistic as well as the diffusive space-time scalings for the two regimes has been shown in Fig.~\ref{fig:correlation_scaling}. A similar calculation can also be done for space-time correlations of other conserved densities, using Eq.~\eqref{eq:correlation_normalised} and Eq.~\eqref{eq:ultimate_reveal}. The steps are straightforward and are not outlined in the paper. 

In summary, we revisited the system of backscattering hard rods in this work. Through the equations of hydrodynamics for fluctuating PSDs, we computed unequal space-time correlations of the normal modes of the problem and found that it takes a diffusive form at large times. A similar computation of the mass densities with the Boltzmann distribution as background state shows that it follow a ballistic space-time scaling initially which then eventually becomes diffusive for late times. It should be stressed that the scope of this formulation is, however, not just restricted to backscattering noise. One may use it to study 
Brownian hard rods to understand the effect of inter-particle interactions in their hydrodynamics. An interesting follow up to this work is to analyze what the transport properties would be if the system of backscattering hard rods were to be confined inside a harmonic trap. 
Such extensions would provide insights relevant to experimental realizations of interacting particle systems confined in a trap and also test the robustness of the framework of Dean-Kawasaki fluctuating hydrodynamics.
\\[7pt]
The author thanks Anupam Kundu and Samriddhi Sankar Ray for insightful discussions and comments on the manuscript. We acknowledge support of the Department of Atomic Energy, Government of India, under project no. RTI4013 and RTI4019 .

\bibliography{references}
\newpage

\onecolumngrid
\vspace{1cm}
\begin{center}
    \textbf{Supplementary Material}
\end{center}

\appendix

\section{Conserved densities of backscattering hard rods} \label{app:conserved_densities}
In this section, we outline the process to compute mean phase space densities in a system of hard rods. Due to the absence of jump events in quasiparticle trajectories, a system of hard PPs is effectively non-interacting. The PSDs in the HR picture are then conveniently expressed in terms of 
PSDs, $\bar{f}(x,v,t)$ in the PP picture by the following
\begin{equation}\label{eq:PSDrod_PSDPP}
\begin{split}
    \bar{\mf{f}}_\sigma(\mt{x},w,t) = \sum_{i=1}^N \binom{N-1}{i-1} \bigg(\frac{\bar{F} (x(\mt{x}_i))}{N}\bigg)^{i-1} \bar{f}_\sigma(x(\mt{x}_i),w,t) \\ \bigg(1-\frac{\bar{F} (x(\mt{x}_i))}{N}\bigg)^{N-i}  
\end{split}
\end{equation}
where, $\bar{F}(x,t) = \int^x \da y  ~\da w~ (\bar{f}_+(y,w,t) + \bar{f}_-(y,w,t))$ is the mean cumulative density upto position $x$ in the PP picture. 

The mapping defined in Eq.~\eqref{eq:rod_to_pp_map} maps the system of BHRs
to a system of non-interacting run and tumble particles (RTPs). The mean PSD for RTPs, $\bar{f}_\sigma(x,\vert v \vert,t)$ satisfies the following Fokker-Planck equation
\cite{malakar2018steady}:
\begin{equation}\label{eq:PP_Fokker-Planck}
\begin{split}
    \frac{\partial \bar{f}_{\sigma} (x,|v|,t)}{\partial t} = &-\sigma \frac{\partial}{\partial x} \bigg( \vert v \vert \bar{f}_{\sigma} (x,|v|,t) \bigg) \\[2pt] + &\gamma (\bar{f}_{-\sigma} (x,|v|,t) - \bar{f}_\sigma (x,|v|,t)).
\end{split}
\end{equation} 
Eq.~\eqref{eq:PP_Fokker-Planck} can be solved for various boundary conditions. For PSDs conditioned 
on fixed initial conditions, i.e., $f_\sigma(x,t|x_0,\vert v \vert)$, we outline the method to solve
Eq.~\eqref{eq:PP_Fokker-Planck} in Appendix~\ref{app:hard_walls}
for hard wall boundary conditions and in Appendix~\ref{app:infinite_line} for an infinite line. By integrating $\bar{f}_\sigma(x,t|x_0,\vert v \vert)$ over initial conditions, one can compute the full
joint PSD, $\bar{f}_{\sigma} (x,|v|,t)$ of RTPs for a particular $\sigma$. When this is feeded into Eq.~\eqref{eq:PSDrod_PSDPP}, it gives us the 
mean PSD, $\mf{f}_\sigma (\mt{x},v,t)$ in the HR picture. 

The conserved densities of a system are computed from the PSDs in 
the following way:
\begin{equation}\label{eq:conserved_den_def}
q_\alpha (\mt{x},t) = \int \text{d}w ~ w^\alpha~ (\mf{f}_+(\mt{x},w,t) + (-1)^\alpha\mf{f}_-(\mt{x},w,t) ) \ \  . 
\end{equation}
Here, $\alpha = 0,1,2$ corresponds to mass, momentum and energy density respectively. In Fig.~\ref{fig:gaussian_massden_enerden}, we show that the expression for the conserved densities defined 
in Eq.~\eqref{eq:conserved_den_def} and obtained using Eq.~\eqref{eq:PSDrod_PSDPP} has an excellent agreement with results from molecular dynamics
simulations. For showing this agreement, the following initial distribution of positions and speeds is considered in the
PP picture:
\begin{equation}\label{eq:initial_cond_gaussian}
    \mu(x_0, \vert v \vert) = \frac{1}{\sqrt{2 \pi \sigma_x^2}} \exp\bigg[- \frac{x_0^2}{2 \sigma_x^2} \bigg] \times \frac{2}{\sqrt{2 \pi T}} \exp\bigg[- \frac{v^2}{2 T} \bigg] \ ,
\end{equation} 
where, $T$ is the temperature. 

\section{Solution of the Fokker-Planck equation for RTPs within hard walls}\label{app:hard_walls}
In this section, we solve the Fokker-Planck equation (Eq.~\eqref{eq:PP_Fokker-Planck}) for RTPs in a box with fully 
reflecting walls. We define the total PSD, $f(x,t|x_0,|v|)$ and the 
current $q(x,t|x_0,|v|)$ conditioned on initial position and speed being $x_0$ and $|v|$ respectively as follows:
\begin{subequations}
    \begin{equation}
        f(x,t|x_0,|v|) = f_+(x,t|x_0,|v|)  + f_-(x,t|x_0,|v|) \ .
    \end{equation}
    \begin{equation}
        q(x,t|x_0,|v|) =  f_+(x,t|x_0,|v|)  - f_-(x,t|x_0,|v|) \ .
    \end{equation}
\end{subequations}
For simplicity of notation through the derivation, we suppress the information about 
the dependence on $x_0$ and $\vert v \vert$ in $f$ and $q$. 
Now, the equations of motion for $f$ and $q$ are given by
\begin{subequations}
    \begin{equation}\label{eq:~dqdt}
        \frac{\partial q(x,t)}{\partial t} = - v \frac{\partial f(x,t)}{\partial x} - 2 \gamma q(x,t)
    \end{equation}
    \begin{equation}\label{eq:~dpdt}
        \frac{\partial f(x,t)}{\partial t} = -v \frac{\partial q(x,t)}{\partial x}
    \end{equation}
\end{subequations}
Upon differentiating Eq.~\eqref{eq:~dpdt} w.r.t time, $t$ and using Eq.~\eqref{eq:~dqdt}, we get
\begin{equation}\label{eq:telegrapher's}
    \frac{\partial^2 f(x,t)}{\partial t^2} + 2 \gamma \frac{\partial f(x,t)}{\partial t} = v^2 \frac{\partial^2 f(x,t)}{\partial x^2} \ .
\end{equation}
The above equation is known as the telegrapher's equation. Let us now state the boundary conditions. 
The RTPs are confined in a one-dimensional box of length $L$, so $x \in [0,L]$. Let us assume that initially,
\begin{equation}
    f_+ (x,t=0) = f_- (x,t=0) = \frac{\delta (x-x_0)}{2} \ .
\end{equation}
This implies that
\begin{equation}
   f(x,t=0) = \delta(x-x_0) \  \ ,  \  \ q(x,t=0) = 0 \ .
\end{equation}
On the boundary, since the walls are perfectly reflecting, we should have
\begin{equation}
    q(x=0, t) = q(x=L, t) = 0 \ .
\end{equation}
Let us now express $f(x,t)$ as the following series:
\begin{equation}
f(x,t) = \sum_{n = -\infty}^{\infty} \mathcal{X}_n(x) \mathcal{T}_n(t) \ ,
\end{equation}
where $\mathcal{X}_n (x)$ ($\mathcal{T}_n(t)$) is a purely space (time) dependent part. Now, when we put this solution in the telegrapher's equation \eqref{eq:telegrapher's}, we get the following equation:
\begin{equation}
\frac{1}{\mathcal{T}_n} \frac{1}{v^2} \frac{\partial^2 \mathcal{T}_n}{\partial t^2} + \frac{2\gamma}{v^2} \frac{1}{\mathcal{T}_n} \frac{\partial \mathcal{T}_n}{\partial t} = \frac{1}{\mathcal{X}_n} \frac{\partial^2 \mathcal{X}_n(x)}{\partial x^2} = - k_n^2 \ \text{(say)} \ ,
\end{equation}
where $k_n$ is a constant.
Let the $n^{\text{th}}$ mode of the space dependent part be given by:
\begin{equation}\label{spaceansatz}
\mathcal{X}_n (x) = a_n \cos (k_n x) + b_n \sin (k_n x) \ ,
\end{equation}
and let the $n^{\text{th}}$ ($n \neq 0$) mode of the time dependent part be given by:
\begin{equation}
\mathcal{T}_n (t) = \exp (-\gamma t) \bigg[c_n \cos (\gamma_n t) + d_n \sin (\gamma_n t) \bigg] \ ,
\end{equation}
where $\gamma_n$ is some constant. We can absorb the constant $c_n$ into the constants $a_n$ and $b_n$ and define $g_n = d_n/c_n$. Thus, 
\begin{equation}\label{timeansatz}
\mathcal{T}_n (t) = \exp (-\gamma t) \bigg[ \cos (\gamma_n t) + g_n \sin (\gamma_n t) \bigg] \\[7pt]
\end{equation}
\paragraph{Boundary conditions:}
For all time, $t$, we want $\frac{\partial f(x,t)}{\partial x}$ to be zero at $x = 0$ and $x = L$. This means that
 \begin{subequations}
\begin{equation}
b_n = 0 \  \forall \ n \ ,
\end{equation}
\begin{equation}
k_n = \frac{n \pi}{L} \ . \\[7pt]
\end{equation}
\end{subequations} 
\paragraph{Initial conditions:}
At time, $t = 0$, we have $\frac{\partial p(x,t)}{\partial t} = 0$ from \eqref{eq:~dpdt}. So, 
\begin{equation}
g_n = \frac{\gamma}{\gamma_n} \ .
\end{equation}
At time, $t = 0$, $p(x, t= 0) = \delta (x - x_0)$. So,
\begin{equation}
\delta (x - x_0) = a_0 + 2 \sum_{n = 1}^\infty a_n  \cos \bigg( \frac{n \pi x}{L} \bigg) \ .
\end{equation}
The coefficients $a_0$ and $a_n$ are hence given by:
\begin{subequations}
\begin{equation}
a_0 = \frac{1}{L} \ ,
\end{equation}
\begin{equation}
a_n = \frac{1}{L} \cos\bigg( \frac{n \pi x_0}{L} \bigg) \ . \\[7pt]
\end{equation}
\end{subequations}
\paragraph{Relation between $\gamma_n$, $\gamma$ and $\beta_n$:}
For non-zero $n$, we have the following equation for $\mathcal{T}_n (t)$:
\begin{equation}\label{Tntele}
\frac{\text{d}^2 \mathcal{T}_n}{\text{d} t^2} + 2 \gamma \frac{\text{d} \mathcal{T}_n}{\text{d} t} + \beta_n \mathcal{T}_n = 0 \ ,
\end{equation}
We now substitute the form of $\mc{T}_n$ from Eq.~\eqref{timeansatz} in the above equation. After working out the derivatives, we get the following relation:
\begin{equation}
\gamma_n^2 = \beta_n - \gamma^2 \ , 
\end{equation}
where
\begin{equation}
\beta_n = \frac{n^2 \pi^2 v^2}{{L}^2} \ . \\[16pt]
\end{equation}
So, the probability distribution $f(x,t)$ is given by
\begin{equation}\label{eq:propagator_parcl_box}
\begin{split}
f(x,t \vert x_0, |v|) = \frac{1}{L} + \frac{2}{L} \sum_{n = 1}^\infty &\cos \bigg( \frac{n \pi x_0}{L} \bigg) \cos \bigg( \frac{n \pi x}{L}  \bigg) \\ &\exp[-\gamma t] \bigg[ \cos \bigg( \sqrt{\beta_n - \gamma^2} t \bigg) + \frac{\gamma}{\sqrt{\beta_n - \gamma^2}} \sin \bigg( \sqrt{\beta_n - \gamma^2} t \bigg) \bigg]
\end{split}
\ ,
\end{equation}
where 
\begin{equation}
\beta_n = \frac{n^2 \pi^2 v^2}{{L}^2} \ .
\end{equation}
To find the expression for $q(x,t)$, we put this expression for $f(x,t)$ in Eq.~\eqref{eq:~dqdt}, and by solving the resulting equation, we get
\begin{equation}
    q(x,t \vert x_0 , |v|) = \bigg(\frac{2}{L}\bigg)\sum_{n=1}^\infty \bigg( \frac{n \pi v}{L} \bigg) \cos \bigg(\frac{n \pi x_0}{L} \bigg) \sin \bigg(\frac{n \pi x}{L} \bigg) \frac{\exp [-\gamma t]}{\sqrt{\beta_n - \gamma^2}} \sin \bigg(\sqrt{\beta_n - \gamma^2} t\bigg) 
\end{equation}

\section{Solution of the Fokker-Planck equation for RTPs on an infinite line}\label{app:infinite_line}
To find the expression on an infinite line, we rewrite Eq.~\eqref{eq:propagator_parcl_box} inside $[-L/2,L/2]$,
\begin{equation}
\begin{split}
f(x,t \vert x_0, |v|) =   &\frac{1}{L} \sum_{n = -\infty}^\infty \cos \bigg( \frac{n \pi (x_0+L/2)}{L} \bigg) \cos \bigg( \frac{n \pi (x + L/2)}{L'}  \bigg) \\ \exp[-\gamma t] &\bigg[ \cos \bigg( \sqrt{\beta_n - \gamma^2} t \bigg) + \frac{\gamma}{\sqrt{\beta_n - \gamma^2}} \sin \bigg( \sqrt{\beta_n - \gamma^2} t \bigg) \bigg]
\end{split}
\ ,
\end{equation}
We define $z = x + L/2$, and change the summation over $n$ to an integral over $k = n\pi/L$, and this becomes
\begin{equation}\label{eq:IFT_form_of_finite_L_expression}
\begin{split}
f(z,t \vert z_0, |v|) = &\frac{1}{2\pi} \int_{-\infty}^\infty \da k ~\exp[i k z] ~(\exp [i k z_0] + \exp [-i k z_0]) ~\exp[-\gamma t] ~ \\ &\bigg[ \cosh (t \sqrt{\gamma^2 - k^2 v^2}) + \frac{\gamma}{\sqrt{\gamma^2 - k^2 v^2}} \sinh (t \sqrt{\gamma^2 - k^2 v^2})   \bigg] \ \ \ .
\end{split}
\end{equation}
One must note that since $x,x_0 > -L/2$, $z,z_0 > 0$ and $x-x_0 = z - z_0$.
The above integration over $k$ is accurate in the limit of large $L$. Since, Eq.\eqref{eq:IFT_form_of_finite_L_expression} is just an inverse fourier transform of a product of two fourier transforms, the following identities would help:
\begin{subequations}
\begin{equation}\label{eq:FT_of_I0}
\int_{-|v|t}^{|v|t} \da y \ I_0 \bigg(\frac{\gamma}{|v|} \sqrt{v^2 t^2 - y^2} \bigg) \exp[-iky] = 2 \frac{\sinh \bigg( |v|t \sqrt{\frac{\gamma^2}{v^2} - k^2} \bigg)}{\sqrt{\frac{\gamma^2}{v^2} - k^2}} \ \ ,
\end{equation}
\begin{equation}\label{eq:FT_of_I1}
\int_{-|v|t}^{|v|t} \da y \frac{\gamma t}{\sqrt{v^2t^2 - y^2}}\ I_1 \bigg(\frac{\gamma}{|v|} \sqrt{v^2t^2 - y^2}\bigg) \exp[-iky]  + 2 \cos (k |v|t) = 2 \cosh \bigg[|v|t \sqrt{\frac{\gamma^2}{v^2} - k^2}\bigg] \ \ ,
\end{equation}
\end{subequations}
where Eq.\eqref{eq:FT_of_I1} is obtained from Eq.\eqref{eq:FT_of_I0} by differentiating both sides w.r.t. $|v|t$. \\[3pt]
The inverse fourier transform of $\exp[ikz_0]$ is $\delta (z+z_0)$. Since, both $z$ and $z_0$ are positive, this solution can be discarded. Hence, we ultimately end up with the following propagator
\begin{equation} \label{eq:prob_sol_fixed_initial}
    \begin{split}
    f(x,t \vert x_0,\vert v \vert) &= \frac{e^{- \gamma t}}{2} \bigg[ \delta(x- \vert v \vert t-x_0) + \delta(x+ \vert v \vert t-x_0) \bigg] \\
    +& \frac{\gamma t e^{-\gamma t}\Theta(\vert v \vert t - \vert x-x_0\vert)}{2} \bigg[\frac{I_1 (\frac{\gamma}{\vert v \vert} \sqrt{v^2 t^2 - (x-x_0)^2})}{\sqrt{v^2 t^2 - (x-x_0)^2}} \\ &+ \frac{1}{\vert v \vert t} I_0 (\frac{\gamma}{\vert v \vert} \sqrt{v^2 t^2 - (x-x_0)^2}) \bigg] \ .
\end{split}
\end{equation}
The same process can be applied for $q(x,t \vert x_0, \vert v\vert)$ and the result gives
\begin{equation}\label{eq:qrob_sol_fixed_initial}
\begin{split}
q (x,t \vert x_0, \vert v \vert) &= \frac{1}{2} e^{-\gamma t} ~ I_1 (\frac{\gamma}{\vert v \vert} \sqrt{v^2 t^2 - (x-x_0)^2}) \frac{\gamma}{\vert v \vert} \frac{\vert x - x_0 \vert \Theta(\vert v \vert t - \vert x-x_0\vert)}{\sqrt{v^2 t^2 - (x-x_0)^2}} \\ &+ \frac{1}{2} e^{-\gamma t} \delta (\vert v \vert t - (x-x_0)) ~ \Theta (x-x_0) \\ &- \frac{1}{2} e^{-\gamma t} \delta (\vert v \vert t - (x_0-x)) ~ \Theta (x_0-x)
\end{split}
\end{equation}
Using Eq.~\eqref{eq:prob_sol_fixed_initial} and Eq.~\eqref{eq:qrob_sol_fixed_initial}, we get for $x>x_0$
\begin{subequations}
\begin{equation}
\begin{split}
f_+ (x,t \vert x_0, \vert v \vert) &= \frac{1}{2} e^{-\gamma t} \delta (\vert v \vert t - (x-x_0)) \\ + &\Theta(\vert v \vert t - \vert x-x_0\vert) \bigg[  \frac{\gamma e^{-\gamma t}}{4 \vert v \vert} I_0 (\frac{\gamma}{\vert v \vert} \sqrt{v^2 t^2 - (x-x_0)^2})  \\ &+ \frac{\gamma e^{-\gamma t}}{4 \vert v \vert} \sqrt{\frac{\vert v \vert t + (x-x_0)}{\vert v \vert t - (x-x_0)}} I_1 (\frac{\gamma}{\vert v \vert} \sqrt{v^2 t^2 - (x-x_0)^2}) \bigg]
\end{split}
\end{equation}
\begin{equation}
\begin{split}
f_- (x,t \vert x_0, \vert v \vert) &= \Theta(\vert v \vert t - \vert x-x_0\vert) \bigg[  \frac{\gamma e^{-\gamma t}}{4 \vert v \vert} I_0 (\frac{\gamma}{\vert v \vert} \sqrt{v^2 t^2 - (x-x_0)^2})  \\ &+ \frac{\gamma e^{-\gamma t}}{4 \vert v \vert} \sqrt{\frac{\vert v \vert t - (x-x_0)}{\vert v \vert t + (x-x_0)}} I_1 (\frac{\gamma}{\vert v \vert} \sqrt{v^2 t^2 - (x-x_0)^2}) \bigg]
\end{split}
\end{equation}
\end{subequations}
while for $x < x_0$, we get
\begin{subequations}
\begin{equation}
\begin{split}
f_+ (x,t \vert x_0, \vert v \vert) &= \Theta(\vert v \vert t - \vert x-x_0\vert) \bigg[  \frac{\gamma e^{-\gamma t}}{4 \vert v \vert} I_0 (\frac{\gamma}{\vert v \vert} \sqrt{v^2 t^2 - (x-x_0)^2})  \\ &+ \frac{\gamma e^{-\gamma t}}{4 \vert v \vert} \sqrt{\frac{\vert v \vert t - (x_0-x)}{\vert v \vert t + (x_0-x)}} I_1 (\frac{\gamma}{\vert v \vert} \sqrt{v^2 t^2 - (x-x_0)^2}) \bigg]
\end{split}
\end{equation}
\begin{equation}
\begin{split}
f_- (x,t \vert x_0, \vert v \vert) &= \frac{1}{2} e^{-\gamma t} \delta (\vert v \vert t - (x_0-x)) \\ + &\Theta(\vert v \vert t - \vert x-x_0\vert) \bigg[  \frac{\gamma e^{-\gamma t}}{4 \vert v \vert} I_0 (\frac{\gamma}{\vert v \vert} \sqrt{v^2 t^2 - (x-x_0)^2})  \\ &+ \frac{\gamma e^{-\gamma t}}{4 \vert v \vert} \sqrt{\frac{\vert v \vert t + (x_0-x)}{\vert v \vert t - (x_0-x)}} I_1 (\frac{\gamma}{\vert v \vert} \sqrt{v^2 t^2 - (x-x_0)^2}) \bigg]
\end{split}
\end{equation}
\end{subequations} 

\section{Full Derivation of fluctuating hydrodynamics}\label{app:derivation_fluctuating_hydro}
To derive the hydrodynamics of this system, we start by writing down the position 
of the $i^\text{th}$ backscattering hard rod (BHR) at time, $t$ by the inverse 
mapping of Eq.~\eqref{eq:rod_to_pp_map} given by
\begin{equation}
    \mt{x}_i (t) = x(\mt{x}_i(t)) + a ~ \sum_{j \neq i} \Theta (x(\mt{x}_i(t)) - x(\mt{x}_j (t))) \ \ , 
\end{equation}
where, $\Theta(x)$ is the heaviside theta function such that $\Theta(x) = 1 \ \text{for} \ x\geq0$ 
and 0 otherwise. Here, we would be tracking the position of the BHR from the position of the 
corresponding PP, which as already stated, is an RTP. In a time interval $\da t$, an RTP moves 
ballistically with its original velocity until a flipping event occurs which changes the direction
of propagation, leading to the combined displacement being partly from the ballistic motion with the
original velocity and rest with the flipped velocity. 
In other words, the full displacement of an RTP in a time interval, $\da t$ is the sum of a pure 
ballistic motion through the entire duration, $\da t$ and a noisy component brought about by the
tumbling motion. Hence, resultant position of the BHR is given by
\begin{equation}\label{eq:HR_pos_tplusdt}
\begin{split}
    \mt{x}_i (t + \da t) = x(\mt{x}_i(t))+ v_i ~\sigma_i \da t + \da \mt{b}_i + &a ~ \sum_{j \neq i} \Theta (x(\mt{x}_i(t)) \\  + v_i ~\sigma_i \da t &+ \da \mt{b}_i - x(\mt{x}_j (t)) - v_j ~\sigma_j \da t - \da \mt{b}_j) \ \ ,
\end{split}
\end{equation}
where,
\begin{equation}
    \da \mt{b}_i = \Delta_i (\da t) ~ \da n_t \ \ \ \ \ , \text{with} \ \ \ \Delta_i (dt) = \mt{x}^-_i (t + \da t) - \mt{x}^+_i (t + \da t) 
\end{equation}
is the noisy component. Here, $\da n_t$ is a random number which is 1 when the particle 
flips its velocity and 0 when it doesn't, $\mt{x}^-_i (t + \da t)$ denotes the position of the 
$i^\text{th}$ hard rod at time, $t+\da t$ had it flipped during the time interval $\da t$ 
and $\mt{x}^+_i (t + \da t)$ the position had it not flipped. For convenience, we can consider 
a $\da t$ during which a maximum of only 1 flipping event occurs.

Before discussing the properties of $\da n_t$, it is important to discuss the limit of $\Delta_i (\da t)$ as $\da t \to 0$. Let us consider a time-interval $\da t' \leq \da t$ during which the particle travels ballistically with its velocity at time, $t$. If a flipping event were to happen ($\da n_t = 1$), it will happen at $\da t'$. Hence, $\Delta_i (\da t)$ is given by (for a speed $w_i$),
\begin{equation}
    \Delta_i (\da t) = w_i \sigma_i \da t' - w_i \sigma_i (\da t - \da t') - w_i \sigma_i \da t = 2 w_i \sigma_i (\da t - \da t') \ .
\end{equation}
Now, at the limit $\da t \to 0$, $\da t' \to \da t$. And hence, $\Delta_i (\da t)$ reaches the value 0, before $\da t$ reaches 0. This can be written in the following way:
\begin{equation}\label{eq:limit_dt}
    \lim_{\da t \to 0} \frac{\Delta_i (\da t)}{\da t} \to 0
\end{equation}

What are the properties of $\da n_t$ ? Since a flipping event is a Poisson process, the probability of 
no flipping event ($\da n_t=0$) is $e^{-\gamma \da t}$ and the probability of exactly one 
flipping event ($\da n_t=1$) is $\gamma \da t e^{-\gamma \da t}$. So, the first moment,
second moment and the second cumulant of $\da n_t$ can be easily
computed:
\begin{equation}
\begin{split}
    \langle \da n_t \rangle = \langle \da n^2_t \rangle  &= 0 (e^{-\gamma \da t})  + 1 (\gamma \da t e^{-\gamma \da t}) = \gamma \da t \ e^{-\gamma \da t} \sim \gamma \da t \ \ \  , \\[2pt]
&\langle \da n^2_t \rangle_c  = \langle \da n^2_t \rangle -  \langle \da n_t \rangle^2 \sim \gamma \da t  .  
\end{split}
\end{equation}

Now, for convenience, we can define a zero mean random number, 
$\tilde{\da n}_t = \da n_t - \gamma \da t$, such that the noise $\da \mt{b}_i$ is
given by 
\begin{equation}
    \da \mt{b}_i = \Delta_i (\da t) ~ \gamma \da t + \Delta_i (\da t) ~ \tilde{\da n}_t \ , \\[3pt]
\end{equation}
where $\langle \tilde{\da n}_t \rangle = 0$ and $\langle \tilde{\da n}^2_t \rangle = \gamma \da t = \langle \tilde{\da n}^2_t \rangle_c$. \\[3pt]

Now we perform a Taylor expansion of
the Heaviside Theta function in Eq.~\eqref{eq:HR_pos_tplusdt} and write down the
change in position of the $i^\text{th}$ BHR as
\begin{equation}
\begin{split}
    \da \mt{x}_i (t) = w_i \sigma_i \da t + \da \mt{b}_i + a \sum_{j \neq i} \delta (x(\mt{x}_i (t)) - x(\mt{x}_j (t))) (w_i \sigma_i \da t + \da \mt{b}_i - w_j \sigma_j \da t - \da \mt{b}_j) \\
    + \frac{a}{2} \sum_{j \neq i} \delta' (x(\mt{x}_i (t)) - x(\mt{x}_j (t)))~ (w_i \sigma_i \da t + \da \mt{b}_i - w_j\sigma_j \da t - \da \mt{b}_j)^2
\end{split}
\end{equation}

We are interested only in terms that are linear in $\da t$ and hence we can
write $\da \mt{x}_i$ as
\begin{equation}
\da \mt{x}_i (t) = \langle \da \mt{x}_i (t) \rangle + \delta \mt{x}_i(t)
\end{equation}
where
\begin{equation}\label{eq:meandXi}
    \langle \da \mt{x}_i (t) \rangle = w_i \sigma_i (t) \da t + a \da t \sum_{j \neq i} \langle \delta ( x(\mt{x}_i (t)) - x(\mt{x}_j (t))) ~ (w_i \sigma_i (t) - w_j \sigma_j (t)) \rangle \ .
\end{equation}
and
\begin{equation}\label{eq:deltaXi}
    \delta \mt{x}_i (t) = \Delta_i (\da t) \tilde{\da n}_t + a \sum_{j \neq i} \delta ( x(\mt{x}_i (t)) - x(\mt{x}_j (t))) ~ (\Delta_i (\da t)~ \tilde{\da n}^{(i)}_t - \Delta_j (\da t)~ \tilde{\da n}^{(j)}_t) \ .
\end{equation}

Any function $\mt{q}$ along the trajectory of the $i^{th}$ BHR can be defined from the fluctuating PSD, $\mf{f}^{(i)}_\sigma (\mt{x}, w, t)$ corresponding to the $i^\text{th}$ BHR i.e.
\begin{equation}
\begin{split}
    \mt{q}_{\sigma_i (t)} (\mt{x}_i(t), w_i (t)) = \sum_\sigma \int \da \mt{x} ~ \da w ~ \mt{q}_\sigma(\mt{x},w) ~ \mf{f}^{(i)}_\sigma (\mt{x}, w, t) \ \ ,
\end{split}
\end{equation}
where $\mf{f}^{(i)}_\sigma (\mt{x}, w, t)$ is as defined in Eq.~\eqref{eq:mainfluctuating_PSD}.
Taking derivative on both sides with respect to time, $t$ and summing over $i$ on both sides gives us
\begin{equation}\label{eq:time_derivative_operator}
\begin{split}
    \frac{\da \mt{Q}}{\da t} &= \sum_\sigma \int \da \mt{x} ~ \da w ~ \mt{q}_\sigma(\mt{x},w) ~ \partial_t \mf{f}_\sigma (\mt{x}, w, t) 
\end{split}
\end{equation}
where we have used Eq. \eqref{eq:mainfluctuating_PSD} and the following definition
of $\mt{Q}(t)$,
\begin{equation}\label{eq:def_q}
\mt{Q} (t) = \sum_i \mt{q}_{\sigma_i (t)} (\mt{x}_i(t), w_i (t)) \ .
\end{equation} 
As can be seen, $\mt{Q}(t)$ is a functional of the trajectories of all the BHRs. 
Now, any change in $\mt{Q}$ in a time interval $\da t$ is given by (to order $\da \mt{x}_i^2$),
\begin{equation} \label{eq:mainchange_in_Q}
\begin{split}
    \da \mt{Q}(t) = \sum_i \frac{\partial \mt{q}_{\sigma_i (t)}(\mt{x}_i(t), w_i (t))}{\partial \mt{x}_i (t)} \da x_i(t) \\[3pt] + \sum_i \da \mt{N}^{(i)}_{\da t} + \mc{O}(\da \mt{x}_i^2) \ .
\end{split}
\end{equation}
Here, the first term corresponds to a change in $\mt{Q}$ brought about by the displacement of each of the BHRs while the second term, $\sum_i \da \mt{N}^{(i)}_{\da t}$ is a noise term due to a flipping event of the BHRs in a time interval $\da t$. 
For Brownian hard rods, the contribution to $\da \mt{Q}$ from terms of $\mc{O}(\da \mt{x}_i^2)$ is also $\mc{O}(\da t)$ and hence should be taken into account.

Now, Eq.~\eqref{eq:mainchange_in_Q} can be rewritten as,
\begin{equation} \label{eq:change_in_Q}
    \da \mt{Q}(t) = \sum_i \frac{\partial \mt{q}_{\sigma_i (t)}(\mt{x}_i(t), w_i (t))}{\partial \mt{x}_i (t)} (\langle \da \mt{x}_i(t) \rangle + \delta \mt{x}_i) + \sum_i \da \mt{N}^{(i)}_{\da t} + \mc{O}(\da \mt{x}_i^2) \ .
\end{equation}
Here, $\da \mt{N}^{(i)}_{\da t}$ is a noise component similar to $\da \mt{b}_i$ defined before
for individual BHRs and is given by
\begin{equation}
\begin{split}
&\da \mt{N}^{(i)}_{\da t} = \Delta_i \mt{q} (\da t) \da n_{\da t} = \Delta_i \mt{q} (\da t)  (\tilde{\da n_{\da t}} + \gamma \da t) \\[3pt]
\text{where} \ \ \ &\Delta_i \mt{q} (\da t) = (\mt{q}^{-}_{\sigma_i(t)} (\mt{x}_i (t + \da t, w_i)) - \mt{q}^{+}_{\sigma_i(t)} (\mt{x}_i (t + \da t, w_i)))
\end{split}
\end{equation}
Now, we use Eq.~\eqref{eq:limit_dt} and the fact that for the test function, as 
$\da t \to 0$, $\Delta_i \mt{q} (\da t) \not \to 0$. This is true, for example, for test functions, $\mt{q}_\sigma$ which are odd in $\sigma$.
Dividing both sides of Eq.~\eqref{eq:change_in_Q} by $\da t$ and then take the 
limit of $\da t \to 0$. In this limit, it is easy to see that $\delta X_i / \da t \to 0$.
Also, by property of Dirac delta function, we know that
\begin{equation}
 \delta(x(\mt{x}_i) - x(\mt{x}_j)) = \frac{\delta(\mt{x}_i - \mt{x}_j)}{1 - a \sum_k \delta(\mt{x}_j - \mt{x}_k)}
\end{equation}
So, Eq.~\eqref{eq:meandXi} gives us
\begin{equation}
\frac{\langle  d \mt{x}_i (t) \rangle }{dt} = \frac{w_i \sigma_i (t) - a \sum_\sigma \int dw ~ w ~\sigma~ \mf{f}_\sigma (\mt{x}_i,w,t)}{1-a \rho (\mt{x}_i,t)}
\end{equation}
where
\begin{equation}
\rho(\mt{x},t) = \sum_j \langle \delta(\mt{x} -\mt{x}_j (t)) \rangle = \sum_j \sum_{\sigma} \int \da w~ \langle \mf{f}^{(j)}_\sigma (\mt{x},w,t) \rangle = \sum_{\sigma} \int \da w~ \bar{\mf{f}}_\sigma (\mt{x},w,t)\ ,
\end{equation}
is the mean mass density. Inserting delta functions in Eq.~\eqref{eq:change_in_Q} and using definitions of fluctuating PSD as in Eq.\eqref{eq:mainfluctuating_PSD}, we get
\begin{equation} \label{eq:change_in_Q_part2}
\begin{split}
    \frac{\da \mt{Q}}{\da t} = &\sum_\sigma \int \da \mt{x}~ \int \da w~ \mf{f}_\sigma (\mt{x},w,t) \frac{\partial \mt{q}_{\sigma}(\mt{x}, w)}{\partial \mt{x}} \frac{\langle \da \mt{x}(t) \rangle}{\da t}  \\  &+ \gamma \sum_\sigma \int \da \mt{x} \int \da w~ (\mf{f}_{-{\sigma}} (\mt{x},w,t) - \mf{f}_{\sigma} (\mt{x},w,t)) \mt{q}_\sigma (\mt{x},w)  \\ 
    &+ \sum_\sigma \int \da \mt{x} \int \da w~ \mt{q}_\sigma (\mt{x},w) (\mf{f}_{-{\sigma}} (\mt{x},w,t) - \mf{f}_{\sigma} (\mt{x},w,t)) \frac{\tilde{\da n}_t}{\da t} .
\end{split}
\end{equation}
Comparing Eqs. \eqref{eq:time_derivative_operator} and \eqref{eq:change_in_Q_part2}, we get
\begin{equation}
\begin{split}
    \partial_t \mf{f}_\sigma (\mt{x},w,t) + \partial_\mt{x} {v_\sigma}_{\rm eff} (\mt{x},w,t) \mf{f}_\sigma (\mt{x},w,t) = \gamma (\mf{f}_{-\sigma} (\mt{x},w,t) - \mf{f}_{\sigma} (\mt{x},w,t)) \\[3pt] +  (\mf{f}_{-\sigma} (\mt{x},w,t) - \mf{f}_{\sigma} (\mt{x},w,t)) \frac{\tilde{\da n}_t}{\da t}
\end{split} 
\end{equation}
where,
\begin{equation}
    {v_\sigma}_{\rm eff} (\mt{x},w,t) = \frac{\langle \da \mt{x} \rangle}{\da t} = \frac{w \sigma - a \sum_{\sigma'} \int \da \mr{w} ~ \mr{w} \sigma' \mf{f}_{\sigma'} (\mt{x},\mr{w},t)}{1-a \rho (\mt{x},t)}.
\end{equation}
Now, define
\begin{equation}
    \xi^{\da t}_\sigma (\mt{x},w,t) = \sum_i (\mf{f}^{(i)}_{-\sigma} (\mt{x},w,t) - \mf{f}^{(i)}_{\sigma} (\mt{x},w,t)) \frac{\tilde{\da n}^{(i)}_t}{\da t} \ .
\end{equation}
The properties of the noise are as follows: 
\begin{equation}
    \langle \xi^{\da t}_\sigma (\mt{x},w,t) \rangle = 0
\end{equation}
and 
\begin{equation}
\begin{split}
    \langle \xi^{\da t}_\sigma (\mt{x},w,t) & ~\xi^{\da t'}_{\sigma'} (\mt{x}',w',t') \rangle_c  \\ =  \bigg\langle \bigg(\sum_i (\mf{f}^{(i)}_{-\sigma} (\mt{x},w,t) - \mf{f}^{(i)}_{\sigma} (\mt{x},w,t)) \frac{\tilde{\da n}^{(i)}_t}{\da t} \bigg) &\bigg(\sum_j (\mf{f}^{(j)}_{-\sigma'} (\mt{x}',w',t') - \mf{f}^{(j)}_{\sigma'} (\mt{x}',w',t')) \frac{\tilde{\da n}^{(j)}_{t'}}{\da t} \bigg)  \bigg\rangle_c \\ 
    = \frac{\gamma}{\da t} \bar{\mf{f}}(\mt{x},w,t) (\delta_{\sigma, \sigma'}-\delta_{\sigma, -\sigma'})~\delta(\mt{x}&-\mt{x}') \delta(w-w') \delta(t-t') \ ,
\end{split}
\end{equation}
where $\bar{\mf{f}}$ is the mean PSD. We can absorb $\sqrt{\gamma \bar{\mf{f}}(\mt{x},w,t)}$ into the definition of the $\xi (\mt{x},w,t)$ and using this new definition we write the equations of fluctuating hydrodynamics as:
\begin{equation}\label{eq:fluc_hydrodynamics}
\begin{split}
    \partial_t \mf{f}_\sigma (\mt{x},w,t) + \partial_\mt{x} {v_\sigma}_{\rm eff} (\mt{x},w,t) \mf{f}_\sigma (\mt{x},w,t) = \gamma (\mf{f}_{-\sigma} (\mt{x},w,t) - \mf{f}_{\sigma} (\mt{x},w,t)) \\[3pt] +  \sigma \sqrt{\gamma~ \bar{\mf{f}}(\mt{x},w,t)} ~ \xi(\mt{x},w,t)
\end{split} 
\end{equation}
with the noise $\xi(\mt{x},w,t)$ having zero mean and having the following property:
\begin{equation}
        \langle \xi (\mt{x},w,t) ~\xi (\mt{x}',w',t') \rangle_c =  \delta(\mt{x}-\mt{x}') \delta(w-w') \delta(t-t')  
\end{equation}

\section{Linearizing equations of fluctuating hydrodynamics}
The linearization of Eq.~\eqref{eq:mainfluc_hydrodynamics}, hence, leads to the following:
\begin{equation}\label{eq:linear_order_fluc_hydro_unnormalized}
\begin{split}
    &\partial_t \delta \mf{f}_\sigma (\mt{x},w,t) + \frac{w \sigma}{1 - a \varrho_0} \partial_\mt{x} \delta \mf{f}_\sigma (\mt{x},w,t) \\ + &\frac{\bar{\mf{f}}_0 (w)}{2} \bigg[\frac{a w \sigma}{(1- a \varrho_0)^2} \partial_\mt{x} \delta \varrho (\mt{x},t) + \frac{\partial_t \delta \varrho (\mt{x},t)}{1-a\varrho_0}  \bigg]  \\ = &\gamma ( \delta \mf{f}_{-\sigma}(\mt{x},w,t) -  \delta \mf{f}_\sigma (\mt{x},w,t)) + \sigma \sqrt{\gamma \bar{\mf{f}}_0} ~\xi (\mt{x},w,t)  \ .
\end{split}
\end{equation}
where, $\delta \varrho(\mt{x},t) = \sum_\sigma \int \da w ~\delta \mf{f}_\sigma (\mt{x},w,t)$ is the resultant fluctuation in the mass density and obeys the following equation of motion:
\begin{equation}
\begin{split}
    \partial_t \delta \varrho(\mt{x},t) + \partial_\mt{x} \delta (\varrho u (\mt{x},t)) = 0 \ \ , \  \ \text{where} \\
    \delta (\varrho u (\mt{x},t)) = \sum_\sigma \int \da w ~ w \sigma ~ \delta \mf{f}_\sigma (\mt{x},w,t) \ \ .
\end{split}
\end{equation}
Eq.~\eqref{eq:linear_order_fluc_hydro_unnormalized} can be simplified by rewriting it in terms of the normal mode fluctuating PSD, $\mf{f}^n_\sigma$ defined as
\begin{equation}\label{eq:def_normalised_PSD}
    \mf{f}^n_\sigma (\mt{x},w,t) = \frac{\mf{f}_\sigma (\mt{x},w,t)}{1-a\varrho(\mt{x},t)} \ .
\end{equation}
Whenever required, we would sometimes refer to $\mf{f}_\sigma$ as the `physical' fluctuating PSD. In terms of $\mf{f}^n_\sigma$, Eq.~\eqref{eq:linear_order_fluc_hydro_unnormalized} can be rewritten as
\begin{equation} \label{eq:linear_order_fluc_hydro}
\begin{split}
    &\partial_t \delta \mf{f}^n_\sigma (\mt{x},w,t) + \frac{w \sigma}{1 - a \varrho_0} \partial_\mt{x} \delta \mf{f}^n_\sigma (\mt{x},w,t) \\ = &\gamma ( \delta \mf{f}^n_{-\sigma}(\mt{x},w,t) -  \delta \mf{f}^n_\sigma (\mt{x},w,t)) + \sigma \sqrt{\gamma \bar{\mf{f}}_0} \frac{~\xi (\mt{x},w,t)}{1 - a \varrho_0} \ .
\end{split}
\end{equation}

\section{Solution to the linearized fluctuating hydrodynamics} \label{app:linearized_fluc}
To solve Eq.~\eqref{eq:linear_order_fluc_hydro} exactly, we need to go to the fourier domain.
Denoting, $\delta \tilde{\mf{f}}^n_\sigma (k,w,t)$ as the fourier transform of $\delta \mf{f}^n_\sigma (\mt{x},w,t)$ with respect to $\mt{x}$, we write the fluctuating hydrodynamics in fourier space as
\begin{equation}\label{eq:fourier_linear_order_fluc_hydro}
    \partial_t \delta \tilde{\mf{f}}^n_\sigma + i \bar{w} \sigma  ~k~ \tilde{\mf{f}}^n_\sigma = \gamma ( \tilde{\mf{f}}^n_{-\sigma} -  \tilde{\mf{f}}^n_\sigma) + \sigma \sqrt{\gamma \bar{\mf{f}}_0} ~ \frac{\tilde{\xi}}{1 - a \varrho_0} \ ,
\end{equation}
where we have suppressed the functional dependence on $k,w$ and $t$ from $\delta \tilde{\mf{f}}^n_\sigma$ and $\tilde{\xi}$ and have defined
\begin{equation}
    \bar{w} = \frac{w}{1 - a \varrho_0} \ .
\end{equation}
We can rewrite the above equation in a matrix notation:
\begin{equation}\label{eq:matrix-perturbation}
    \partial_t ~\widetilde{\delta F^n} (k,w,t) + A(k,w)~ \widetilde{\delta F^n} (k,w,t) =  \sqrt{\gamma \bar{\mf{f}}_0} ~ \tilde{\Xi^n} (k,w,t)
\end{equation}
where,
\begin{equation}
    \widetilde{\delta F} = (\tilde{\delta f^n_+} \ \ \ \ \tilde{\delta f^n_-})^T  \ \ \ \ , \ \  \ \ \ \tilde{\Xi} = \frac{1}{1 - a\varrho_0} (\tilde{\xi} \ \ \ \ \  \\ - \tilde{\xi})^T
\end{equation}
and 
\begin{equation}
    A = \begin{pmatrix}
        i k \bar{w} + \gamma & -\gamma \\
        -\gamma & -ik \bar{w} + \gamma
    \end{pmatrix} \ . 
\end{equation}
The solution to \eqref{eq:matrix-perturbation} is given by
\begin{equation}
    \widetilde{\delta F} (k,w,t) = \exp [- A t]~ \widetilde{\delta F} (k,w,0) + \sqrt{\gamma \bar{\mf{f}}_0} \int_0^t dt' ~ \exp[-A(t-t')] ~\Xi(k,w,t')
\end{equation}
where $\exp[-A t]$ is given by
\begin{equation}\label{eq:Solution_in_matrix_form}
    \begin{pmatrix}
        e^{-\gamma t} \bigg( \cosh[t \sqrt{\gamma^2 - k^2 \bar{w}^2}] - \frac{i k \bar{w} \sinh[t \sqrt{\gamma^2 - k^2 \bar{w}^2}]}{\sqrt{\gamma^2 - k^2 \bar{w}^2}} \bigg) & \frac{\gamma e^{-\gamma t} \sinh[t \sqrt{\gamma^2 - k^2 \bar{w}^2}]}{\sqrt{\gamma^2 - k^2 \bar{w}^2}} \\
        \frac{\gamma e^{-\gamma t} \sinh[t \sqrt{\gamma^2 - k^2 \bar{w}^2}]}{\sqrt{\gamma^2 - k^2 \bar{w}^2}} & e^{-\gamma t} \bigg( \cosh[t \sqrt{\gamma^2 - k^2 \bar{w}^2}] + \frac{i k \bar{w} \sinh[t \sqrt{\gamma^2 - k^2 \bar{w}^2}]}{\sqrt{\gamma^2 - k^2 \bar{w}^2}} \bigg)
    \end{pmatrix} \ .
\end{equation}

Let us define the following
\begin{subequations}
    \begin{equation}
    \begin{split}
        \mt{S}_{++} (k,w,t) = &\cosh[t \sqrt{\gamma^2 - k^2 \bar{w}^2}] - i k \bar{w}~ \frac{\sinh[t \sqrt{\gamma^2 - k^2 \bar{w}^2}]}{\sqrt{\gamma^2 - k^2 \bar{w}^2}}  
    \end{split}
    \end{equation}
    \begin{equation}
    \mt{S}_{+-} (k,w,t) = \frac{\gamma  \sinh[t \sqrt{\gamma^2 - k^2 \bar{w}^2}]}{\sqrt{\gamma^2 - k^2 \bar{w}^2}}  
    \end{equation}
\end{subequations}

The solution in fourier space is then given by
\begin{subequations}\label{eq:solution_in_component_form}
\begin{equation}
    \delta \tilde{\mf{f}}^n_+ (t) = e^{-\gamma t} \bigg(~\mt{S}_{++}(t)~ \delta\tilde{\mf{f}}^n_+ (0) + ~\mt{S}_{+-}(t)~ \delta\tilde{\mf{f}}^n_- (0) \bigg) + \frac{\sqrt{\gamma \bar{\mf{f}}_0}}{1 - a \varrho_0} \int_0^t \da t' ~e^{-\gamma(t-t')}~\tilde{\xi}(t')\bigg(~\mt{S}_{++}(t-t') - \mt{S}_{+-}(t-t') \bigg) \ ,
\end{equation}
\begin{equation}
    \delta \tilde{\mf{f}}^n_- (t) = e^{-\gamma t} \bigg(~\mt{S}^\dagger_{++}(t)~ \delta\tilde{\mf{f}}^n_- (0) + ~\mt{S}_{+-}(t)~ \delta\tilde{\mf{f}}^n_+ (0) \bigg) + \frac{\sqrt{\gamma \bar{\mf{f}}_0}}{1 - a \varrho_0} \int_0^t \da t' ~e^{-\gamma(t-t')}~\tilde{\xi}(t')~\bigg(~\mt{S}_{+-}(t-t') - \mt{S}^\dagger_{++}(t-t') \bigg) \ ,
\end{equation}
\end{subequations}
where we have suppressed the dependence on $k$ and $\bar{w}$. The dagger($\dagger$) represents complex conjugate. This solution can then be used to compute space-time correlations of the form:
\begin{equation}\label{eq:def_fourier_transform}
\begin{split}
    &\langle  \delta \mf{f}^n (\mt{x},w,t) ~ \delta \mf{f}^n (\mt{x}',w',t') \rangle \\[3pt] = &\frac{1}{(2 \pi)^2} \int \da k ~\da k' ~ e^{i k \mt{x} +i k' \mt{x}'}~ \langle  \delta \tilde{\mf{f}}^n (k,w,t) ~ \delta \tilde{\mf{f}}^n (k',w',t') \rangle
\end{split}
\end{equation}
where $\delta \mf{f}^n = \delta \mf{f}^n_+ + \delta \mf{f}^n_-$ is the total fluctuating PSD and the tilde's denote fourier transforms. Ignoring the term that depends on the initial value of the fluctuation, and using the properties of the noise as given in and around Eq.\eqref{eq:noise_property_in_fourier_space}
we get (for $t>t'$),
\begin{equation}
\begin{split}
    \langle  \delta \tilde{\mf{f}}^n (k,w,t) &~ \delta \tilde{\mf{f}}^n (k',w',t') \rangle = 2 \pi~ \frac{\gamma ~\bar{\mf{f}}_0 (w)}{(1 - a \varrho_0)^2}  ~ \delta(k+k') ~ \delta(w-w') ~ \exp[-\gamma (t+t')] \bigg[ - \frac{\sinh[(t+t')\sqrt{\gamma^2 - k^2 \bar{w}^2}] }{\sqrt{\gamma^2 - k^2 \bar{w}^2}} \\
    & + \exp[2 \gamma t'] \frac{\sinh[(t-t')\sqrt{\gamma^2 - k^2 \bar{w}^2}] }{\sqrt{\gamma^2 - k^2 \bar{w}^2}} + \frac{\exp[2 \gamma t']}{\gamma} \cosh[(t-t')\sqrt{\gamma^2 - k^2 \bar{w}^2}] \\ 
    & - 2 \gamma \frac{\sinh[t\sqrt{\gamma^2 - k^2 \bar{w}^2}] }{\sqrt{\gamma^2 - k^2 \bar{w}^2}} \frac{\sinh[t'\sqrt{\gamma^2 - k^2 \bar{w}^2}] }{\sqrt{\gamma^2 - k^2 \bar{w}^2}} - \frac{1}{\gamma}\cosh[(t-t')\sqrt{\gamma^2 - k^2 \bar{w}^2}]  \bigg] \ .
\end{split}
\end{equation}
We can easily see that in the limit of $t,t' >>\gamma^{-1}$), we get
\begin{equation}\label{eq:correlation_in_fourier_space}
\begin{split}
    &\langle  \delta \tilde{\mf{f}}^n (k,w,t) ~ \delta \tilde{\mf{f}}^n (k',w',t') \rangle \\ =  &\frac{ 2 \pi~\gamma ~\bar{\mf{f}}_0 (w)}{(1 - a \varrho_0)^2}  ~ \delta(k+k') ~ \delta(w-w') ~ \exp[-\gamma (t-t')] \\
    & \bigg[ \frac{\sinh[(t-t')\sqrt{\gamma^2 - k^2 \bar{w}^2}] }{\sqrt{\gamma^2 - k^2 \bar{w}^2}} + \frac{1}{\gamma} \cosh[(t-t')\sqrt{\gamma^2 - k^2 \bar{w}^2}] \bigg] \ .
\end{split}
\end{equation}
The term inside the integral in Eq.~\eqref{eq:def_fourier_transform} is the correlation function of normal mode PSDs in fourier space and in the limit of $t,t' >>\gamma^{-1}$, is given by Eq.~\eqref{eq:correlation_in_fourier_space}. 
Taking an inverse fourier transform of the same and using the identities given in Eq.~\eqref{eq:FT_of_I0} and Eq.~\eqref{eq:FT_of_I1}, we get the correlation function of normalised PSDs in real space to be
\begin{equation}\label{eq:correlation_normalised}
\begin{split}
    \langle  \delta \mf{f}^n (\mt{x},w,t) ~ \delta \mf{f}^n (\mt{x}',w',t') \rangle  = &\frac{\bar{\mf{f}}_0 (w)}{2 (1-a\varrho_0)^2}  ~\delta(w-w')\exp[-\gamma (t-t')] \\ &\Bigg[\delta (\mt{x}-\mt{x}'+ \bar{w} (t-t')) + \delta (\mt{x}-\mt{x}'-\bar{w} (t-t')) \\
    &+\Bigg(\frac{\gamma}{\bar{w}} I_0 \bigg(\frac{\gamma}{\bar{w}} \sqrt{\bar{w}^2 (t-t')^2 - (\mt{x}-\mt{x}')^2}\bigg)  +  ~ ~ ~ ~ \gamma(t-t') \frac{~ I_1 (\frac{\gamma}{\bar{w}} \sqrt{\bar{w}^2 (t-t')^2 - (\mt{x}-\mt{x}')^2})}{\sqrt{ \bar{w}^2 (t-t')^2 - (\mt{x}-\mt{x}')^2}} \Bigg)  \\ & \ \ \ \ \ \ \ \ \  \Theta (\bar{w} (t-t')  - \vert \mt{x} - \mt{x'}\vert )  \Bigg]  \ .
\end{split}
\end{equation}
From here, one may find the late-time behaviour as shown in Eq.~\eqref{eq:diffusive_late_time_behaviour} by utilizing the following property:
\begin{equation}\label{eq:prop_large_z}
    I_\nu (z) \sim \frac{\exp[z]}{\sqrt{2 \pi z}} \ , \ \ \ \ \  \text{for large $z$} 
\end{equation}
and for any $\nu$.

From the definition of the normal mode PSD,
the variation, $\delta \mf{f}^n $ due to the fluctuation, $\delta \mf{f}$ is given by
\begin{equation}
    \delta \mf{f}^n (\mt{x},w,t) = \frac{\delta \mf{f} (\mt{x},w,t)}{1 - a \varrho_0} + \frac{a \bar{\mf{f}}_0 (w)}{(1-a\varrho_0)^2} \delta \varrho(\mt{x},t) \ .
\end{equation}
Using the above equation,
the space-time correlation of normal mode PSDs is related to that of the physical PSDs, in the following way:
\begin{equation} \label{eq:ultimate_reveal}
\begin{split}
    &\langle  \delta \mf{f}^n (\mt{x},w,t) ~ \delta \mf{f}^n (\mt{x}',w',t') \rangle \\[3pt] = &\frac{\langle \delta \mf{f}(\mt{x},w,t)\delta \mf{f}(\mt{x}',w',t')\rangle}{(1-a\varrho_0)^2}  + \frac{a \bar{\mf{f}}_0(w)}{(1-a\varrho_0)^3} \langle \delta \varrho (\mt{x},t) \delta \mf{f}(\mt{x}',w',t') \rangle 
     \\
     +  &\frac{a \bar{\mf{f}}_0(w')}{(1-a\varrho_0)^3} \langle \delta \mf{f}(\mt{x},w,t) \delta \varrho (\mt{x}',t')\rangle  + \frac{a^2 \bar{\mf{f}}_0(w) \bar{\mf{f}}_0(w')}{(1-a\varrho_0)^4} \langle \delta \varrho(\mt{x},t) \delta \varrho (\mt{x}',t')\rangle \ .
\end{split}
\end{equation}
To find the space-time correlations of mass densities, we integrate both sides with respect to $w$ and $w$'. This gives us
\begin{equation}
\begin{split}
    \langle \delta \varrho^n(\mt{x},t) \delta \varrho^n (\mt{x}',t)\rangle = &\frac{\langle \delta \varrho(\mt{x},t) \delta \varrho (\mt{x}',t)\rangle}{(1-a\varrho_0)^2} + \frac{a \varrho_0 \langle \delta \varrho(\mt{x},t) \delta \varrho (\mt{x}',t)\rangle}{(1-a \varrho_0)^3}  \\
    &+\frac{a \varrho_0 \langle \delta \varrho(\mt{x},t) \delta \varrho (\mt{x}',t)\rangle}{(1-a \varrho_0)^3} + \frac{a^2 \varrho_0^2 \langle \delta \varrho(\mt{x},t) \delta \varrho (\mt{x}',t)\rangle}{(1-a \varrho_0)^4}
\end{split}
\end{equation}
Finally, rearranging the terms, one gets
\begin{equation}\label{eq:correlation_relation_normal_unnormal}
    \langle \delta \varrho^n(\mt{x},t) \delta \varrho^n (\mt{x}',t)\rangle = \frac{\langle \delta \varrho(\mt{x},t) \delta \varrho (\mt{x}',t)\rangle}{(1-a \varrho_0)^4}
\end{equation}
To compute $\langle \delta \varrho^n(\mt{x},t) \delta \varrho^n (\mt{x}',t)\rangle$ for the state given in
Eq.~\eqref{eq:Gibbs_state}, we 
integrate both sides of Eq.~\eqref{eq:correlation_normalised} with respect to $w$ and $w'$. We define $\tau = t-t'$ and 
$\Delta = |\mt{x} - \mt{x}'|$. 

For $\tau <<\gamma^{-1}$, only the delta function term would contribute to a space-time scaling.
The integration of the delta function is straightforward. Let us denote it as $\mc{S}_2$ and hence,
\begin{equation}
    \mathcal{S}_2 (\Delta, \tau) =\frac{\varrho_0}{(1-a\varrho_0)} \frac{e^{-\gamma \tau}}{\sqrt{2 \pi T~ \tau^2}} \exp \bigg[ - (1-a\varrho_0)^2~\frac{\Delta^2}{2 T \tau^2} \bigg] \ . 
\end{equation}

For $\tau >> \gamma^{-1}$, only the bessel functions would contribute. This makes the lower limit of integration over $w$ to be 0.
Let us denote the integration of
Bessel $I_0$ as $\mc{S}_0$,
\begin{equation}\label{eq:integral_over_I0}
    \mathcal{S}_0 (\Delta,\tau) = \frac{e^{-\gamma \tau}}{(1-a\varrho_0)^2} \frac{\gamma \varrho_0}{\sqrt{2 \pi T}} \int_0^\infty \da w \ \frac{1}{\bar{w}} \ \exp\bigg[- \frac{w^2}{2T}\bigg] ~ I_0 \bigg(\gamma \tau \sqrt{1 - \frac{\Delta^2}{\bar{w}^2 \tau^2}} \bigg) 
\end{equation}
Now using the property given in Eq.~\eqref{eq:prop_large_z}, we can write Eq.~\eqref{eq:integral_over_I0} as
\begin{equation}
\begin{split}
    &\mathcal{S}_0 (\Delta,\tau) \\ =& \frac{e^{-\gamma \tau}}{(1-a\varrho_0)^2}\frac{\gamma \varrho_0}{ 2 \pi \sqrt{T \gamma \tau}} \int_0^\infty \da w~ \bigg(1- \frac{\Delta^2}{\bar{w}^2~\tau^2} \bigg)^{-1/4} \ \frac{1}{\bar{w}} \ \exp\bigg[- \frac{w^2}{2T} + \gamma \tau \sqrt{1 - \frac{\Delta^2}{\bar{w}^2~ \tau^2}} \bigg] \\[6pt]
    =& \ \ \ \ \  \frac{e^{-\gamma \tau}}{2\pi} \sqrt{\frac{\gamma}{T \tau}} \frac{\varrho_0}{(1-a\varrho_0)} \int_0^\infty \da w \frac{1}{w} \exp \bigg[- \frac{w^2}{2T} + \gamma \tau \bigg(1-\frac{\Delta^2}{2 \bar{w}^2 \tau^2} \bigg) \bigg] \ \ ,
\end{split}
\end{equation}
where we have used binomial expression, $\sqrt{1 - b} \sim 1 - b/2$ for small $b$ inside the exponential function. Now, by changing the variables to $u = w^{-2}$, we can express the integral above as
\begin{equation}
    \mathcal{S}_0 (\Delta,\tau) = \frac{1}{4\pi} \sqrt{\frac{\gamma}{T \tau}} \frac{\varrho_0}{(1-a\varrho_0)} \int_0^\infty \da u \frac{1}{u} \exp \bigg[ - \frac{1}{2T} \frac{1}{u} - \frac{\gamma \Delta^2 (1-a\varrho_0)^2}{2 \tau} u \bigg] \ .
\end{equation}
The above integral gives the Bessel K of zeroth order. Hence, we write
\begin{equation}
    \mathcal{S}_0 (\Delta,\tau) = \frac{\varrho_0}{2\pi (1-a\varrho_0)} \sqrt{\frac{\gamma}{T\tau}} K_0 \bigg( (1-a \varrho_0) \sqrt{\frac{\gamma}{T \tau}}~\Delta  \bigg) \ .
\end{equation}
The integral of the Bessel I of first order can be similarly performed by considering that for large $t$,
\begin{equation}
    \frac{I_1 (\frac{\gamma}{\bar{w}} \sqrt{\bar{w}^2 \tau^2 - \Delta^2})}{\sqrt{\bar{w}^2 \tau^2 - \Delta^2}} \sim \frac{I_1 \bigg( \gamma \tau \ \sqrt{1 - \frac{\Delta^2}{\bar{w}^2 \tau^2}} \bigg)}{\bar{w} \tau} \ ,
\end{equation}
and using the large $z$ form of $I_\nu (z)$, we get that
\begin{equation}
\begin{split}
    &\mathcal{S}_1 (\Delta,\tau) \\ =& \frac{\varrho_0}{(1-a\varrho_0)^2} \frac{\gamma \tau e^{-\gamma t}}{2t} \frac{2}{\sqrt{2 \pi T}} \int_0^\infty \da w \ \frac{1}{\sqrt{\bar{w}^2 \tau^2 - \Delta^2}} \ \exp\bigg[- \frac{w^2}{2T}\bigg] ~ I_1 \bigg(\gamma \tau \sqrt{1 - \frac{\Delta^2}{\bar{w}^2 \tau^2}} \bigg) \\[5pt]
    =& \mathcal{S}_0 (\Delta,\tau) = \frac{\varrho_0 }{2\pi (1-a\varrho_0)} \sqrt{\frac{\gamma}{T\tau}} K_0 \bigg((1-a\varrho_0)\sqrt{\frac{\gamma}{T \tau}}~\Delta  \bigg)
\end{split}
\end{equation}
Thus, 
\begin{equation}\label{eq:unnormal_corr}
\begin{split}
    &\langle \delta \varrho^n(\mt{x},t) \delta \varrho^n (\mt{x}',t)\rangle =  \\
    =&\frac{\varrho_0}{(1-a\varrho_0)} \frac{e^{-\gamma \tau}}{\sqrt{2 \pi T~ \tau^2}} \exp \bigg[ - (1-a\varrho_0)^2~\frac{(\mt{x}-\mt{x}')^2}{2 T \tau^2} \bigg] \ \text{for } \tau<<\gamma^{-1} \text{ and, }\\[3pt] & \frac{\varrho_0}{\pi (1-a\varrho_0)} \sqrt{\frac{\gamma}{T\tau}} K_0 \bigg[(1-a\varrho_0)\sqrt{\frac{\gamma}{T \tau}}~|\mt{x}-\mt{x}'|  \bigg] \ \text{for } \tau>>\gamma^{-1}.
\end{split}
\end{equation}
One may now use Eq.~\eqref{eq:correlation_relation_normal_unnormal} and Eq.\eqref{eq:unnormal_corr} to obtain Eq.\eqref{eq:unequal_time_space_time}.

\end{document}